\documentstyle[pra,twocolumn,aps,psfig]{revtex}
\def\lsim{\raise0.3ex\hbox{$<$\kern-0.75em\raise-1.1ex\hbox{$\sim$}}}
\begin{document}
\draft
\title{Time-dependent tunneling of Bose-Einstein condensates}
\author{O.~Zobay and B.~M.~Garraway}
\address{Sussex Centre for Optical and Atomic Physics, University of
Sussex, Brighton, BN1 9QH, United Kingdom
\medskip}
\author{\small\parbox{14.2cm}{\small\hspace*{3mm}
The influence of atomic interactions on time-dependent tunneling processes
of Bose-Einstein condensates is investigated. In a variety of contexts
the relevant condensate dynamics can be 
described by a Landau-Zener equation modified by the appearance
of nonlinear contributions. Based on this equation it is discussed how
the interactions modify the tunneling probability.
In particular, it is shown that for certain parameter values, due to a
nonlinear hysteresis effect, complete adiabatic population transfer is
impossible however slowly the resonance is crossed. The results also indicate
that the interactions can cause significant increase as well as decrease of
tunneling probabilities which should be observable in currently feasible
experiments.
\\[3pt]PACS numbers: 03.75.Fi, 03.75.Be, 05.30.Jp, 42.50.Vk}}
\maketitle

\section{Introduction}

Since the first experimental preparation of Bose-Einstein condensates (BECs)
in dilute atomic gases the study of their dynamical properties has
been a very active field of research. It has
become apparent that atomic interactions play a crucial role in the
prediction or explanation of a wide range of observable phenomena
including, e.g., free condensate expansion, collective excitations,
nonlinear atom optics, solitons, and vortices \cite{DalGioPit99,KetDurSta99}.
The recent experiment
of Ref.\ \cite{AndKas98} has drawn attention to a further dynamical
process, namely time-dependent tunneling.
This work investigated the dynamics of a BEC which was accelerated
by gravity in the periodic potential formed by two counterpropagating
vertical laser beams. In this way, Bloch oscillations of the condensate
were induced, and each time a turning point of the oscillation was reached
a fraction of the atoms tunneled into a continuum Bloch band.
The regular output of atom pulses spectacularly proved the macroscopic
coherence of the initially prepared condensate.

In the experiment of Ref.\ \cite{AndKas98} the influence of atomic interactions
mainly showed up as a degradation in the interference when condensates with
high densities were studied. As a further consequence, long-time
dephasing of the pulse output was predicted. However, another interesting
question arising in this context concerns the problem of how
the atomic interactions affect the individual tunneling processes which
are fundamental to the dynamics of the system.
The purpose of this paper is to work out essential aspects of
this question by studying the modification of the tunneling probability
in the single process. To this end, the condensate dynamics is
modeled in terms of a mean-field description provided by the
Gross-Pitaevskii equation. If the periodic potential is sufficiently weak
and its period short compared to the condensate extension this equation can
be simplified by means of a two-mode expansion. Thereby, the two modes
represent the original and
the Bragg-scattered contribution. In this way one arrives at a
set of equations similar to the familiar Landau-Zener problem \cite{Lan32}.
In the present case, however, the equations also contain nonlinear terms
that characterize the effect of the
interactions. It should be emphasized that these equations, which are
central to our investigation, are applicable not only to the
description of Bloch band tunneling but are of broader scope. They can also
model processes such as population
transfer between different hyperfine states by variable external fields
\cite{MewAndKur97} or the motion of BECs on coupled potential surfaces
(which might be of interest in practical applications). These processes,
as well as Bloch band tunneling, are examples of coherent output coupling
from BECs \cite{KozDenHag99}
so that the present work also relates to this context. Furthermore, it
extends recent studies of nonlinear Josephson oscillations that are
based on the time-independent version of these equations
\cite{SmeFanGio97,RagSmeFan99,WilWalCoo99}. The time-independent
equations are relevant not only in the context of Bose condensation
but apply to other areas of physics as well, e.g., polaron dynamics, nonlinear
optics, biophysics, and molecular physics \cite{RagSmeFan99,EilLomSco85}
so that the present study might also be of interest in some of these fields.

The paper is organized as follows. In Sec.\ II the model is set up
and the `nonlinear Landau-Zener equations' mentioned above are derived.
Section III first gives a brief qualitative discussion of how the nonlinearity
affects the tunneling probability. We then discuss the main result of our
investigation which shows that for certain parameter values the
tunneling probability does not vanish however slowly the system evolves.
In other words, a complete adiabatic population transfer is impossible
under such circumstances. This effect which would not be possible in
linear two-level crossing models is a direct consequence of the discrete
self-trapping transition \cite{EilLomSco85}. Using a
phase space representation of the problem the tunneling probability
for very slow processes can be determined. Subsequently, we
discuss the population transfer for fast and slow resonance
crossing using numerical methods and simple analytical models. In Sec.\ IV
the predictions of the two-mode system are compared to the numerical solution
of the Gross-Pitaevskii equation describing Bloch-band tunneling. These
studies also show that significant modifications of the tunneling
probability due to nonlinear effects should be observable in
currently feasible experiments. Conclusions are given in Sec.\ V.

\section{The model}

The Gross-Pitaevskii equation for a condensate wave function $\psi({\bf
r},t)$ undergoing Bloch band tunneling in a periodic optical potential
is given by
\begin{equation} \label{GPE}
i\hbar\dot\psi=H_{lin}\psi +g|\psi|^2\psi
\end{equation}
with \cite{PeiDahBou97,NiuRai98} 
\begin{equation}
H_{lin}=-\frac{\hbar^2\nabla^2}{2m}+(V_T+V_{opt}-Fz).
\end{equation}
The atomic mass is denoted $m$, and $g=4\pi\hbar^2 a N/m$ is the nonlinearity
parameter with $a$ the $s$-wave scattering length and $N$ the
number of atoms in the sample. The condensate is assumed to be tightly
confined in the radial direction by the trapping potential $V_T$, the $z$
dependence of which is neglected. The periodic optical potential is given by
$V_{opt}=V_0\cos(2k_Lz)$, with the laser wave vector $k_L$, whereas the
potential inducing the Bloch oscillations is characterized
by the accelerating force $F$. Such a potential may be produced by
tilting the standing wave \cite{AndKas98} or by frequency-shifting the
counterpropagating laser fields \cite{BhaMadMor97}. We want to discuss
the condensate time evolution under the following two conditions. First, 
the condensate has an initial momentum $p_0$ well defined on the
scale set by the lattice wave vector $2k_L$, i.e., $l_z\gg \lambda=\pi/k_L$
with $l_z$ the axial extension of the BEC. The time evolution starts with the
BEC well separated from any tunneling resonances, i.e., avoided
crossings between Bloch bands.
These resonances occur for condensate momenta around
$(2l+1)\hbar k_L$, $l=0,\pm 1,\dots$ For concreteness we take $p_0=0$
in the following [apart from the example of Fig.\ \ref{Fig5}(b)].
The condensate is studied as it passes through the resonance at
$\hbar k_L$. The second condition is that the optical potential is sufficiently
weak, i.e., $V_0 \ll \hbar^2 k_L^2/2m$. In this case, the crossing of the
resonance can be described in terms of a coupling to a single
state with a momentum shifted by $-2\hbar k_L$. The
wave function $\psi({\bf r},t)$ is expanded in terms of the
original and the Bragg-scattered contribution, i.e.,
\begin{eqnarray} \label{expansion}
\psi&=&e^{i(k_L+Ft/\hbar)z}\phi_+\Big(z-\frac{\hbar k_L}m t
-\frac{F t^2}{2m}\Big) b_+(t)\\
&&+e^{i(-k_L+Ft/\hbar)z}\phi_-\Big(z+\frac{\hbar k_L}m t -\frac{F
t^2}{2m}\Big)b_-(t) \nonumber
\end{eqnarray}
with $t=0$ corresponding to the point of exact resonance and $b_+=1$,
$b_-=0$, initially. The two envelope functions $\phi_{\pm}$ (whose
radial motion is frozen due to the tight confinement) are normalized
to one. It is assumed that $\phi_-$ is very similar to
$\phi_+$, the shift in position between the two being negligible at all
times $t$ relevant for the tunneling process, i.e., we set
$\int d^3r \phi_-(z)\phi_+(z-2\hbar k_L t/m)\approx 1$. This approach is
valid if the nonlinear effects are not too large and the tunneling
process is sufficiently short. Expression (\ref{expansion}) is then
inserted into Eq.\ (\ref{GPE}) thereby discarding second derivatives of
$\phi_{\pm}$ (slowly-varying envelope approximation). By projecting the
resulting equation onto the instantaneous wave functions $e^{i(\pm k_L+Ft/
\hbar)z} \phi_{\pm}$ and switching to a rotating frame where
$a_{\pm}(t)=\exp[i\int^t \omega(t')dt']b_{\pm}(t)$, $\omega(t)=\hbar
k_L^2/2m+F^2 t^2/2 m \hbar-\varepsilon t/2-2 \gamma$,
we obtain the `nonlinear Landau-Zener equations'
\begin{eqnarray}\label{NLZ}
i\dot a_+ & = & \varepsilon t a_+ + \Omega a_- +\gamma |a_+|^2 a_+, \nonumber
\\
i\dot a_- & = & \Omega a_+ +\gamma |a_-|^2 a_-
\end{eqnarray}
with
\begin{equation}
\varepsilon=\frac{2Fk_L}m\ ,\ \Omega=\frac{V_0}{2\hbar}\ ,\ \gamma=-\frac
g {\hbar}\int\! d^3 r\, |\phi({\bf r})|^4.
\end{equation}
In the following, $\Omega > 0$ is assumed. As mentioned in the
Introduction similar equations can be derived in other
contexts, e.g., for population transfer between different
hyperfine ground states with variable external fields \cite{MewAndKur97}.
In Sec.\ IV it is shown that Eqs.\ (\ref{NLZ}) allow an accurate prediction
of tunneling probabilities in realistic situations.

Significant insight into the system behavior can be obtained by noting
that Eqs.\ (\ref{NLZ}) may be derived from the Hamilton function
\cite{SmeFanGio97,EilLomSco85}
\begin{eqnarray}\label{Ham}
H(N_+,\Theta)&=&\Delta N_+ +2\Omega \sqrt{N_+(1-N_+)}\cos\Theta
\nonumber \\
&& + \gamma(N_+^2-N_+ +1/2)
\end{eqnarray}
with $N_+=|a_+|^2$ and $\Theta=\mbox{arg}(a_+a_-^*)$ as canonical
variables and $\Delta=\varepsilon t$. The dynamics induced by this Hamilton
function for fixed, time-independent $\Delta$ is discussed in detail in
Refs.\ \cite{SmeFanGio97,RagSmeFan99,EilLomSco85}. In the present context the
stationary states $S=(N_S,\Theta_S)$ of $H(N_+,\Theta)$ are of particular
interest because they take over the role of the adiabatic eigenstates in
the linear problem ($\gamma=0$). From $\partial H/\partial \Theta=0$
it follows that they always have $\Theta_S=0$ or $\pi$. The condition
$\partial H/\partial N_+=0$ then shows that $N_S=n_s+1/2$ is obtained
as a real-valued solution of the equation
\begin{equation}\label{sstat}
n^4_s+\frac{\Delta}{\gamma}n^3_s+\left(\frac{\Omega^2}{\gamma^2}+\frac{\Delta^2}
{4\gamma^2}-\frac 1 4\right)n^2_s-\frac{\Delta}{4\gamma}n_s -\frac{\Delta^2}
{16\gamma^2}=0.
\end{equation}
If $|\gamma|/\Omega\leq 2$ there are exactly two stationary
states $S_-$ and $S_+$ having $\Theta=0$ and $\pi$, respectively, for all
values of $\Delta$. They correspond to the high- and low-energy
eigenstates of the linear problem.
For $|\gamma|/\Omega > 2$ two further states appear in the
vicinity of $\Delta=0$ as discussed in more detail in Sec.\ III.

\section{Tunneling probabilities for the nonlinear Landau-Zener model}

\newpage

\subsection{Overview and qualitative discussion}

In the study of the tunneling probability for the nonlinear Landau-Zener
model we are interested in the
long-time solution of Eqs.\ (\ref{NLZ}) provided the system is
prepared at $t\to -\infty$ in the high- or low-energy (stationary) state.
We choose the low-energy state $a_+$ in the
following, i.e., $a_+(t\to -\infty)=1$, and the tunneling probability is
thus $P_T=|a_+(t\to \infty)|^2$. In the linear case
($\gamma=0$) the tunneling probability
is independent of whether the system starts in the low- or the
high-energy state. In the nonlinear case the same still holds true
if the sign of the nonlinearity is changed as well, i.e., the tunneling
probability for a system starting in `$+$' is the same as for one
initially in `$-$' and having a nonlinearity parameter $-\gamma$.
The whole range of dynamical phenomena predicted
by Eqs.\ (\ref{NLZ}) (for both signs of $\gamma$) can thus be observed with a
condensate of repulsive interactions by using both initial conditions. Due to
the scaling properties of Eqs.\ (\ref{NLZ}) $P_T$ can be considered a function
of $\Omega/ \varepsilon^2$ and $\gamma/\Omega$, only. An overview of the
dependence of $P_T$ on these parameters is given in Fig.\ \ref{Fig1}
which shows results obtained from the numerical solution of Eqs.\
(\ref{NLZ}). In the linear problem the numerical calculation of $P_T$ is
facilitated by transforming from $a_+,a_-$ onto the
basis of the instantaneous `dressed' eigenstates. In this basis the
oscillations of the system trajectory around its asymptotic limit for
$t\to \infty$ are suppressed and after a relatively short propagation
time the value of $P_T$ can be read off from the population of the eigenstates.
This approach is generalized to the nonlinear case by computing
the canonical action of the trajectories as determined by the Hamilton
function (\ref{Ham}). This action plays the role of the adiabatic
invariant and, in the linear case, is equivalent to the eigenstate population.
Various aspects of the behavior of $P_T$ are discussed below.

For the linear problem the tunneling probability is given exactly by the
well-known Landau-Zener formula $P_T=\exp(-2\pi\Omega^2/\varepsilon)$
\cite{Lan32}. A qualitative understanding of how the nonlinear terms
influence $P_T$ can
be obtained by noting that they give rise to an effective detuning
$\Delta_{eff}=\varepsilon t +\gamma(2|a_+|^2-1)$ between the states
$a_+$ and $a_-$. Unless the resonance is traversed too rapidly the system
remains in the vicinity of $S_+(\varepsilon t)$ and the population $|a_+(t)|^2$
closely follows its steady state value $N_{S,+}=n_{s,+}(\varepsilon
t)+1/2$. A rough estimate of the tunneling probability can be
obtained from the rate $R=\dot\Delta_{eff}(t=0)$
at which the point of
zero detuning is crossed. As a first approximation it follows from the
linear Landau-Zener formula that $P_T\approx \exp(-2\pi\Omega^2/R)$. As for
small $\Delta$
\begin{equation}\label{nsexp}
n_{s,+}\approx -\Delta/[2(2\Omega+\gamma)]+\Omega\Delta^3/[2(2\Omega+\gamma)^4]
\end{equation}
one thus finds
\begin{equation}\label{ptapp}
P_T\approx\exp\{-2\pi\frac{\Omega^2}{\varepsilon}
(1+\gamma/2\Omega)\}.
\end{equation}
\centerline{\psfig{figure=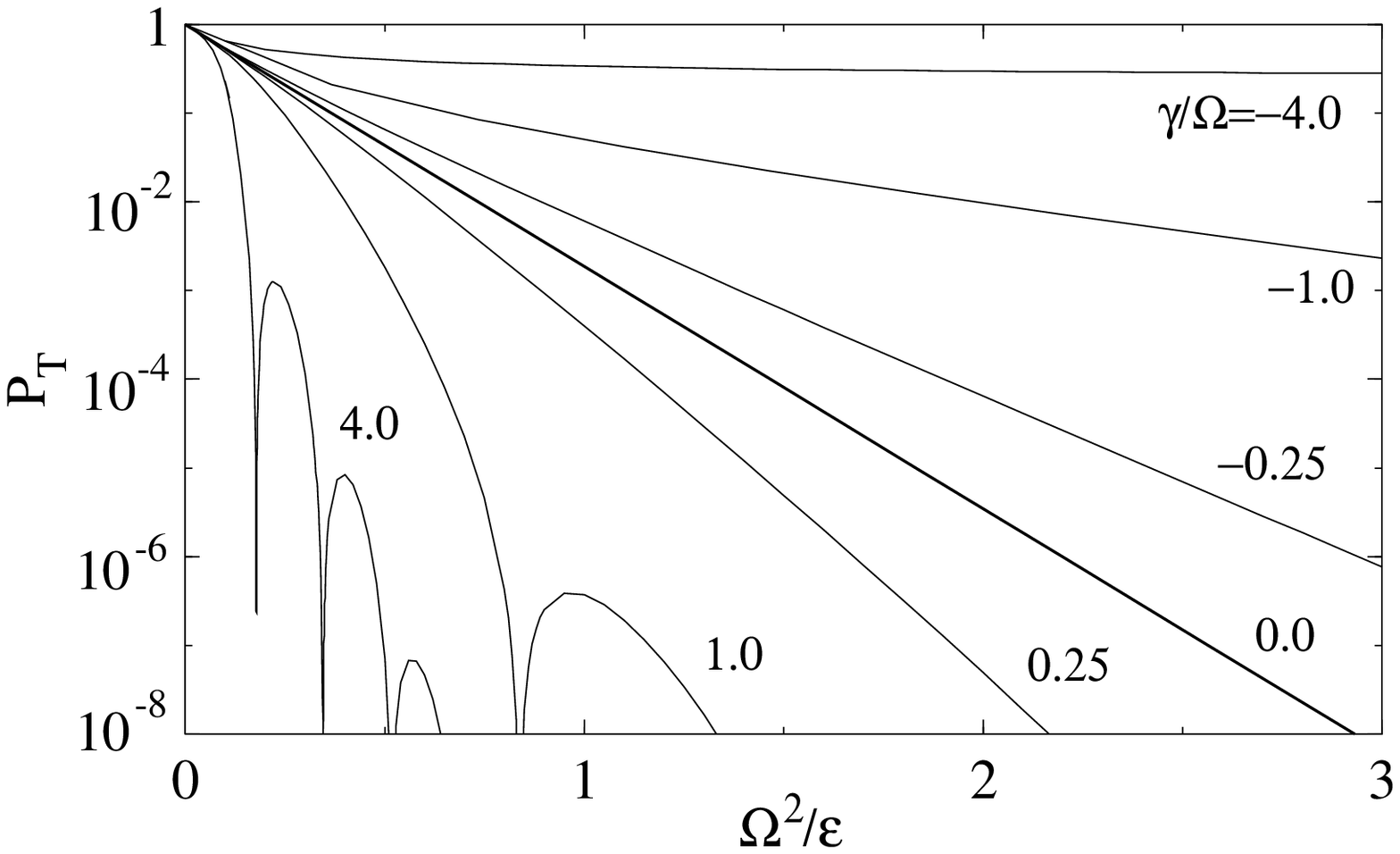,width=8.6cm,clip=}}
\begin{figure}
\caption{Tunneling probability $P_T$ as a function of
$\Omega^2/\varepsilon$ for various values of $\gamma/\Omega$.}
\label{Fig1}
\end{figure}
\noindent
Although not quantitatively accurate this
formula indicates two trends confirmed by the detailed investigation:
(i) nonlinear effects become significant as soon as $\gamma$ becomes
comparable to $\Omega$ (as one might expect) and (ii)
$P_T$ is increased
(decreased) for $\gamma>0$ ($\gamma<0$) as compared to the linear case.
Interestingly, the expansion of $n_{s,+}$ also
shows that the third-order correction to $\Delta_{eff}$ renders the
tunneling transition superlinear (sublinear) for $\gamma>0$ ($\gamma<0$)
in the terminology of Ref.\ \cite{VitSuo99}, i.e., the resonance is
effectively crossed faster (slower) than the linear approximation
predicts \cite{Rem1}. The results of Ref.\ \cite{VitSuo99} indicate
that the approximation (\ref{ptapp}) should
underestimate the nonlinear effects on $P_T$. This is indeed observed in the
present case if $\Omega/\varepsilon^2$ is not too small and $\gamma/\Omega$
sufficiently larger than $-2$.

\subsection{Nonlinear hysteresis effects in the tunneling probability}

The above discussion already indicates the significance of the stationary
state behavior for an understanding of the tunneling probability. The
breakdown of the expansion for $n_{s,+}$ at $\gamma/\Omega=-2$ now
suggests that at this point a qualitative change in the system behavior
occurs. Indeed, for $\gamma/\Omega<-2$, as $\Delta$ is increased from large
negative values, two further stationary states of $H(N_+,\Theta)$,
$S'_+$ and $U_+$, emerge at a detuning $-\Delta_c<0$ [see Fig.\
\ref{Fig2}(a)]. They both have the
same phase $\Theta=\pi$ as $S_+$ and populations $N_{S',+}<N_{U,+}<N_{S,+}$.
The point $S'_+$ is stable whereas $U_+$ is unstable. With growing
$\Delta$, $U_+$ approaches $S_+$. At $\Delta_c$ they
coalesce and only $S'_+$ remains (besides $S_-$). We thus observe a
typical nonlinear hysteresis phenomenon. How does this scenario affect
the tunneling
process? As long as $\gamma/\Omega>-2$, for $\varepsilon \to
0$ the system will stay
arbitrarily close to $S_+$ over the whole evolution of the system so that
$P_T\to 0$. For $\gamma/\Omega<-2$,
however, the hysteresis effect gives rise to a different behavior which
is sketched in Fig.\ \ref{Fig2}(b).
For small $\varepsilon$ the system will again remain
close to $S_+$, initially. However, as $U_+$ approaches
\centerline{\psfig{figure=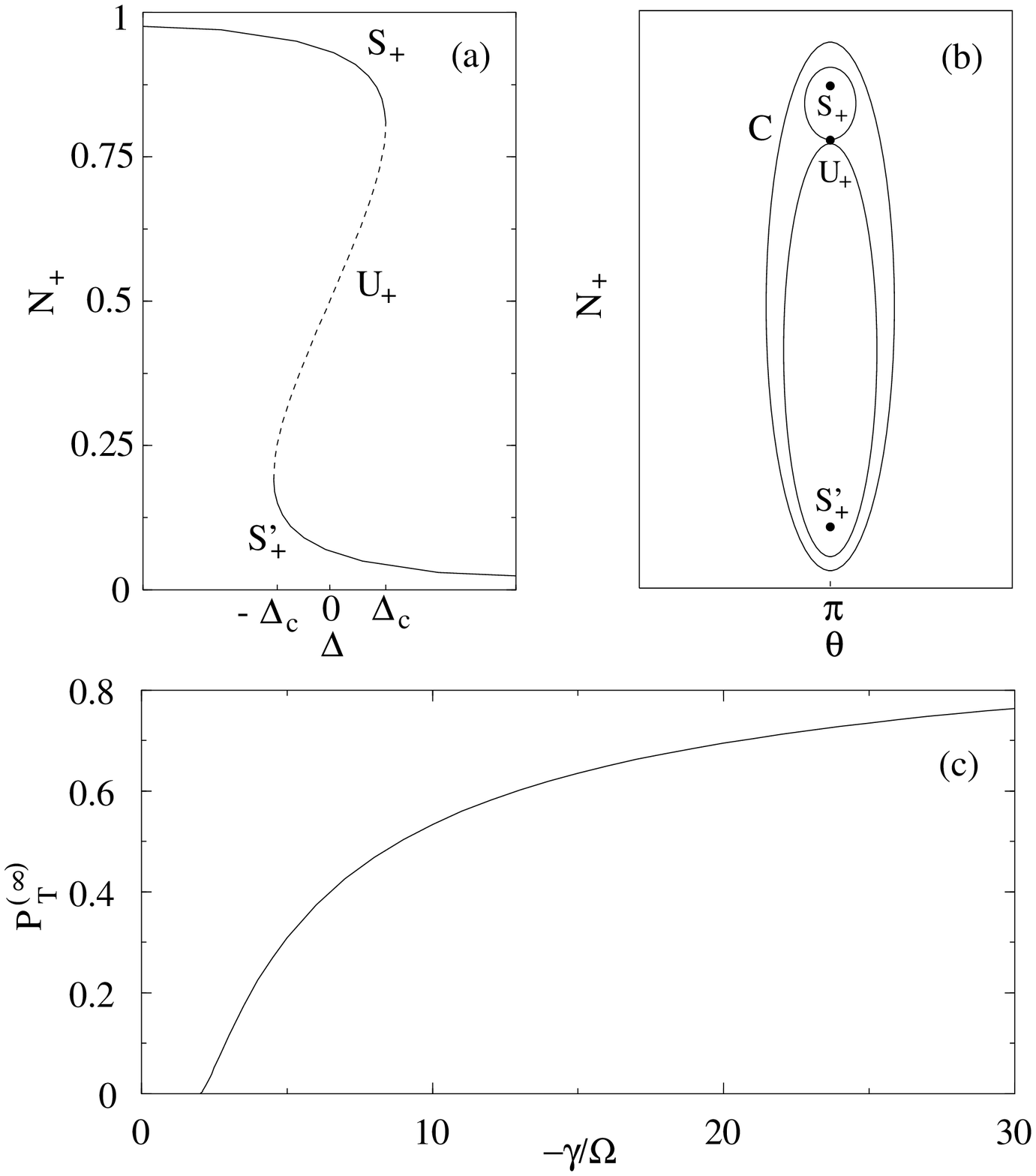,width=8.6cm,clip=}}
\begin{figure}
\caption{(a) Coordinate $N_+$ of stationary states as a funcion of
$\Delta$ for $\gamma/\Omega<-2$. (b) Schematic phase space plot in the
vicinity of $\Delta_c$. The system trajectory eventually has to switch
onto the orbit $C$.
(c) Asymptotic probability $P_T^{(\infty)}$ as a function of $-\gamma/\Omega$.}
\label{Fig2}
\end{figure}
\noindent
$S_+$ it eventually has to switch from its trajectory in the
vicinity of $S_+$ onto a large orbit $C$ encircling $S'_+$ \cite{Rem2}.
Subsequently, the system evolution is
quasi-adiabatic as $t\to \infty$. The final tunneling probability is
thus determined by the canonical action of the orbit $C$ which is an
adiabatic invariant in this case. We thus conclude that in the limit
$\Omega^2/\varepsilon\to \infty$ the tunneling probability is given by
\begin{equation}\label{eps0}
P_T^{(\infty)} = P_T(\Omega^2/\varepsilon\to \infty)=
\frac 1 {2\pi}\oint_{C^{\star}} \Theta\, dN_+
\end{equation}
with $C^{\star}$ the orbit passing through $U_+=S_+$ at $\Delta=\Delta_c$ and
encircling $S'_+$. For small $\varepsilon > 0$ the orbit $C$, onto which
the system switches, completely encloses $C^{\star}$. Therefore,
$P_T^{(\infty)}$ can be expected to be a (at least local) minimum
of the tunneling probability. In Fig.\ \ref{Fig2}(c) $P_T^{(\infty)}$ is shown
as a function of $-\gamma/\Omega$. It becomes apparent
that, even for modest values of $-\gamma/\Omega$, $P_T^{(\infty)}$
is not small compared to one. The above discussion
is confirmed by the numerical
simulation of Eqs.\ (\ref{NLZ}). In particular, it is found that for
fixed $-\gamma/\Omega<-2$, $P_T$ is a monotonically decreasing function of
$\Omega^2/\varepsilon$ (cf.\ Fig.\ \ref{Fig1}).
As $\Omega^2/\varepsilon \to \infty$, it tends to the
value given by Eq.\ (\ref{eps0}) which is thus a global lower bound on the
tunneling probability. Furthermore, for small $\varepsilon$, the canonical
action along the system trajectory abruptly changes around $\Delta_c$, as
expected.

\subsection{Rapid passage through resonance}

Having discussed the asymptotic limit of the tunneling probability we
now study more closely its behavior for small values of $\Omega^2/
\varepsilon$, i.e., when the resonance is passed rapidly. In this case
some insight may be gained from a perturbative analysis of
Eqs.\ (\ref{NLZ}) with $\Omega$ the small
parameter. In a rotating frame with $a_+=\tilde
a_+\exp(-i\varepsilon t^2/2-i\gamma t)$, $a_-=\tilde a_-$ Eqs.\
(\ref{NLZ}) read
\begin{eqnarray}\label{NLZpert}
i\dot{\tilde a}_+ & = & \Omega \tilde a_-\exp(i\varepsilon t^2/2 +
i\gamma t) +\gamma (|\tilde a_+|^2-1) \tilde a_+, \\
i\dot{\tilde a}_- & = & \Omega \tilde a_+\exp(-i\varepsilon t^2/2 - i\gamma t)
 +\gamma |\tilde a_-|^2 \tilde a_-.
\end{eqnarray}
These equations are now iterated in the standard way with initial conditions
$\tilde a^{(0)}_+(-\infty)=1$, $\tilde a^{(0)}_-(-\infty)=0$ up to third
order. This yields
\begin{eqnarray}
&&\tilde a^{(3)}_-(t)=-i\Omega\int_{-\infty}^t\!\! dt_1\, {\cal E}(t_1) \\
&& +i\Omega^3\int_{-\infty}^t\!\! dt_1 \int_{-\infty}^{t_1}\!\! dt_2
\int_{-\infty}^{t_2}\!\! dt_3\, {\cal E}(t_1){\cal E}^*(t_2){\cal E}(t_3)
\nonumber \\
&& -\gamma\Omega^3 \int_{-\infty}^t\!\! dt_1 \Big|\int_{-\infty}^{t_1}\!\! dt_2
{\cal E}(t_2)\Big|^2 \int_{-\infty}^{t_1}\!\! dt_2
\, {\cal E}(t_2)+O(\Omega^5) \nonumber\\
&& \equiv\Omega T_1+\Omega^3 T_3 +O(\Omega^5) \nonumber
\end{eqnarray}
with ${\cal E}(t)=\exp(-i\varepsilon t^2/2 - i\gamma t)$. The first
integral converges for $t\to \infty$ whereas the other two diverge.
This behavior is to be expected because, e.g., in the linear problem
($\gamma=0$), although the moduli of $\tilde a_{\pm}$ converge,
their phases contain logarithmically divergent terms which are reflected
in the divergence of the $\Omega^3$ contributions. A convergent approximation
of $P_T$ up to order $\Omega^4$ can nevertheless be obtained from
$\tilde a^{(3)}_-(t\to\infty)$. Writing $P_T\approx 1-|\tilde a^{(3)}_-
(t\to\infty)|^2=1-\Omega^2|T_1|^2-2\Omega^4\mbox{Re}(T_1^*T_3)+O(\Omega^6)$
it is found that the expression $\mbox{Re}(T_1^*T_3)$ converges; all
divergent contributions in the $\Omega^3$ integrals are contained in 
$\mbox{Im}(T_1^*T_3)$. Determining the limits $t\to\infty$ of the
relevant expressions the approach finally yields
\begin{equation}\label{pert}
P_T=1-2\pi\frac{\Omega^2}{\varepsilon}+2\pi^2\frac{\Omega^4}{\varepsilon^2}
-{\cal N}\frac{\gamma\Omega^4}{\varepsilon^{2.5}}+O(\Omega^6).
\end{equation}
The first three terms are familiar from the expansion of the Landau-Zener
formula $P_T=\exp(-2\pi\Omega^2/\varepsilon)$ for the
linear problem. The fourth term represents the first nonvanishing contribution
of the nonlinearity. The numerical coefficient ${\cal N}$ is equal
to Re$[e^{-i\pi/4}
\sqrt{128\pi}\int_{-\infty}^{\infty} dt |{\cal F}(t)|^2{\cal F}(t)]\approx
15.751$ with ${\cal F}(t)=\int_{-\infty}^t dt_1\exp(-it_1^2)$.

At this point it has to be emphasized that it is not claimed that
Eq.\ (\ref{pert}) is rigorously valid as expansion of $P_T$ around
$\Omega=\gamma=0$. Nevertheless, it yields the following useful
information:
(i) For very rapid passage through resonance the nonlinear interactions
do not influence the tunneling probability. Their contributions arise
\centerline{\psfig{figure=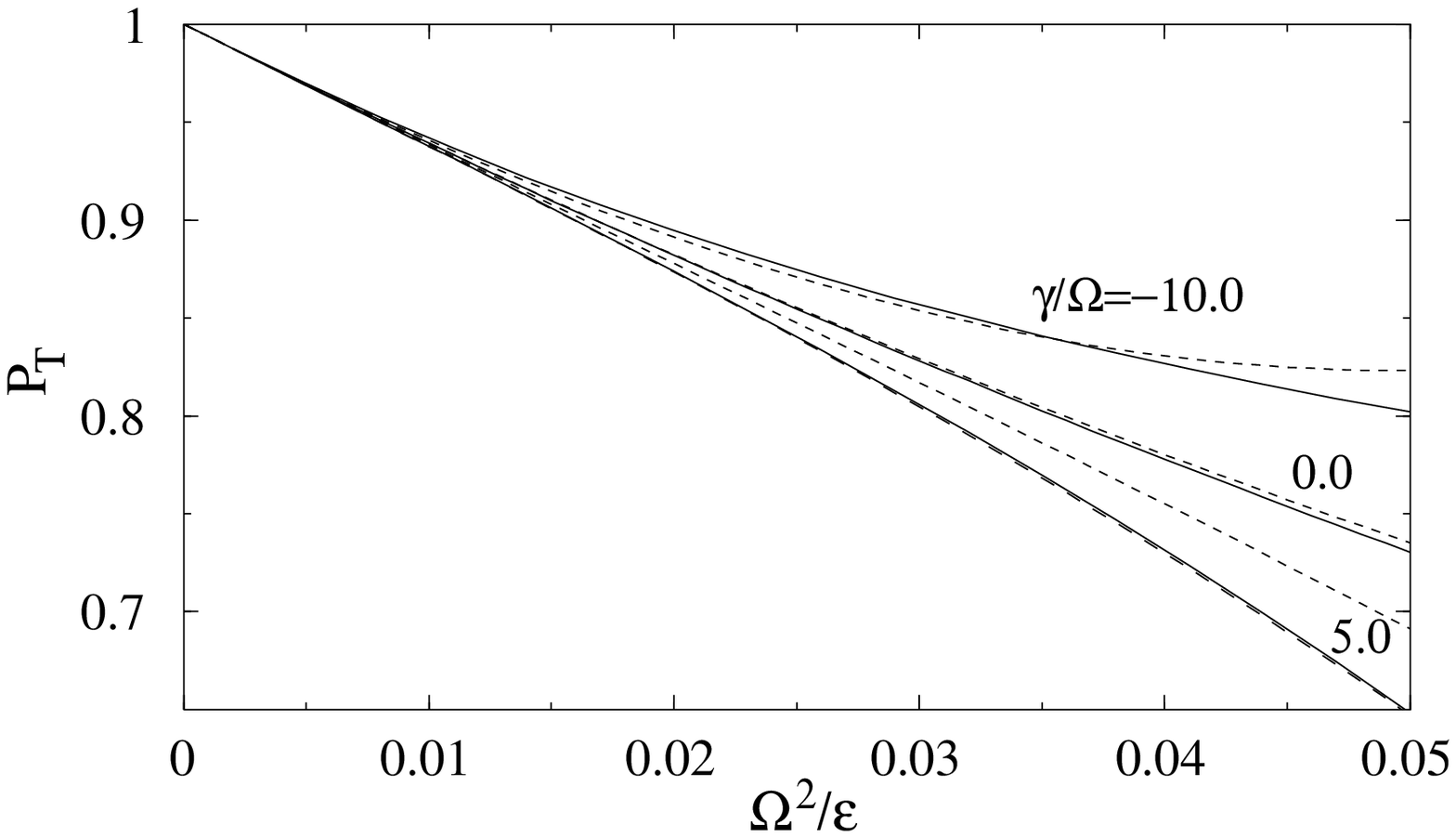,width=8.6cm,clip=}}
\begin{figure}
\caption{Numerical calculation (full curves) of $P_T$  and approximation
according to Eq.\ (\protect\ref{pert}) (dashed curves) for
$\gamma/\Omega=-10.0,0.0,5.0$. The result of Eq.\ (\protect\ref{pert}) for
$\gamma/\Omega=10.0$ is given by the long-dashed curve that is close to
the numerical one for $\gamma/\Omega=5.0$.}
\label{Fig3}
\end{figure}
\noindent
only in higher order in $\Omega$ (see Fig.\ \ref{Fig3}).
(ii) Equation (\ref{pert})
correctly
predicts that for $\gamma>0$ ($\gamma<0)$ $P_T$ is diminshed (increased).
(iii) Quantitatively, Eq.\ (\ref{pert}) gives a good approximation
of $P_T$ in particular for negative $\gamma$ and could be used, e.g., at
$\gamma/\Omega=-10$ to estimate $P_T$ for $\Omega^2/\varepsilon < 0.04$
(cf.\ Fig.\ \ref{Fig3}).
For positive $\gamma$ the agreement is not as good. In the interesting
regime $1<\gamma/\Omega<10$, $\Omega^2/\varepsilon < 0.04$, the
nonlinear effects are underestimated by a factor of about 2. This is 
because the influence of higher-order terms in $\gamma$, which
are neglected in Eq.\ (\ref{pert}), is more relevant for positive $\gamma$. 

\subsection{Slow passage through resonance}

To examine the tunneling probability for slow passage through resonance in
more detail we again use the idea that under these circumstances the system
stays close to the instantaneous stationary state $S_+$. A simple analytically
tractable model is obtained if in Eqs.\ (\ref{NLZ}) $|a_+|^2$ is
replaced by the stationary state population $N_{S,+}(t)$ of the {\it linear}
system, i.e., $|a_+|^2\approx 1/2 - \Delta/[4(\Delta^2/4+\Omega^2)^{1/2}]$
with $\Delta=\varepsilon t$. The tunneling probability $P_T^{(l)}$
for this system can be estimated with the help of the semiclassical theory of
nonadiabatic transitions \cite{VitSuo99,Dyk62,DavPec76,SuoGarSte91} which is
based on the study
of the quasienergy function $Q(t)=[\Delta_{eff}^2(t)/4+\Omega^2]^{1/2}$.
Thereby, $\Delta_{eff}=\varepsilon t +\gamma(2|a_+|^2-1)$. If $t_k$ are the
zeroes (or complex crossing points) of $Q(t)$ in the upper complex half plane
the tunneling probability can be approximated as $P_T^{(l)}=|\sum_k
\Gamma_k \exp(i{\cal D}_k)|^2$ with ${\cal D}_k=2\int_0^{t_k}Q(t)dt$ 
and $\Gamma_k=-2i\lim_{t\to t_k}(t-t_k)\Omega\dot\Delta_{eff}(t)/Q^2(t)$.
As briefly shown in the Appendix,
for $|\gamma|/\Omega \lsim 1$, $P_T^{(l)}$ is well approximated by
\begin{eqnarray}\label{PAD}
P_T^{(l)}&=&\exp(-I') [2\cos(2\pi\gamma\Omega/3\varepsilon)-1]^2,
\ \gamma>0, \nonumber\\
P_T^{(l)}&=&\exp(-I'),\ \gamma<0,
\end{eqnarray}
\centerline{\psfig{figure=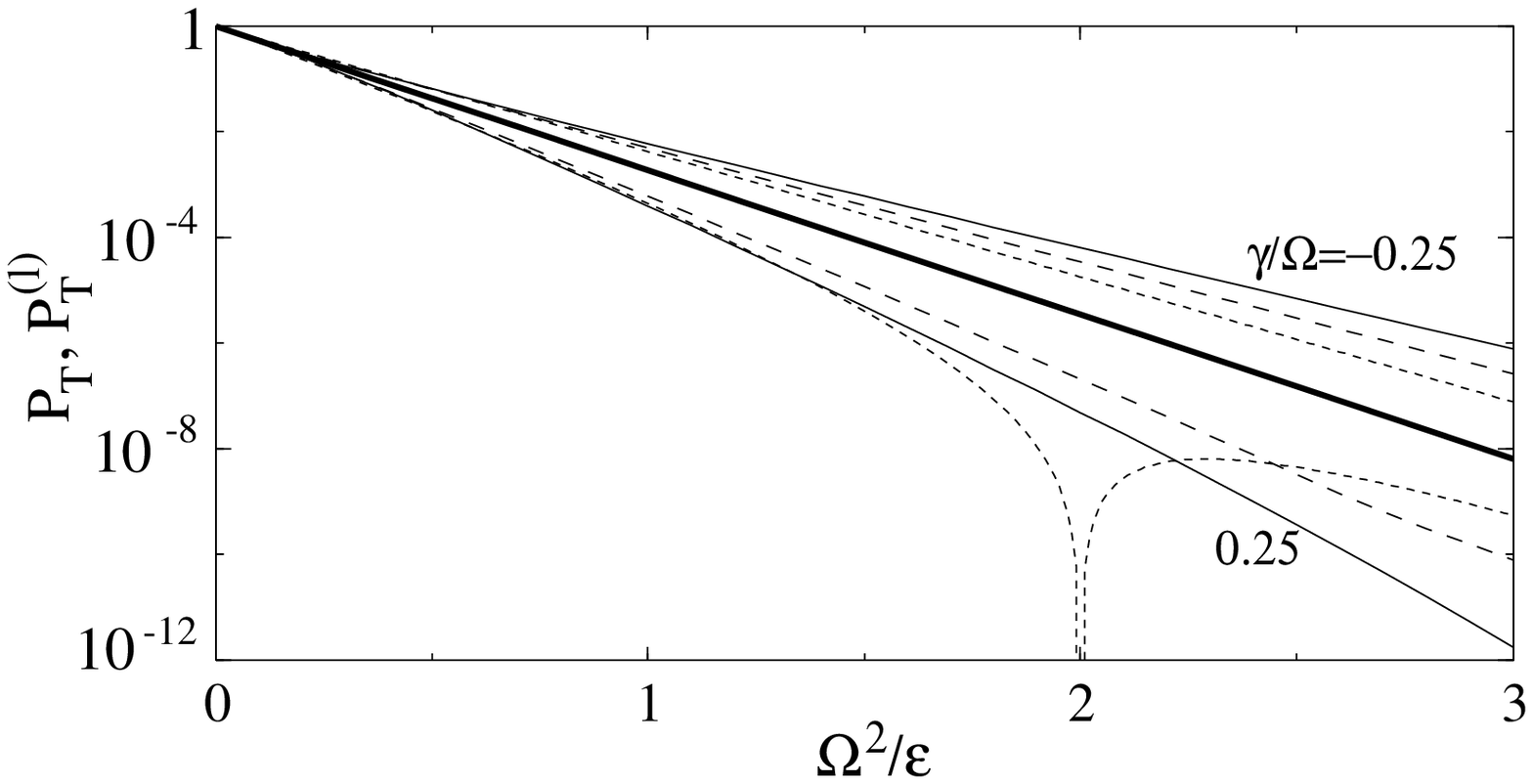,width=8.6cm,clip=}}
\begin{figure}
\caption{Tunneling probability for $\gamma/\Omega=\pm 0.25$. Bold
curves: exact results obtained from Eqs.\ (\protect\ref{NLZ}),
short-dashed curves: approximation according to Eqs.\ (\protect\ref{PAD}).
The long-dashed curves show the tunneling probability with $|a_+|^2$
in Eqs.\ (\protect\ref{NLZ}) replaced by the exact stationary state
population as determined from Eq.\ (\protect\ref{sstat}).
For larger $|\gamma|/\Omega$ the discrepancy is even more pronounced.
The thick line shows $P_T$ for $\gamma=0$.}
\label{Fig4}
\end{figure}
with $I'=2\Omega^2\{\pi-\gamma[\frac 2 3 +2\log(\sqrt[3]{|\gamma|/\Omega}/2)]\}
/\varepsilon$.
Comparison to the solution of Eqs.\ (\ref{NLZ})
shows that the formula for $\gamma>0$ gives a useful approximation
to $P_T$ well before the first minimum of $P_T^{(l)}$, i.e., if
$\Omega^2/\varepsilon \lsim \Omega/4\gamma$ (see Fig.\ \ref{Fig4}).
For negative $\gamma$
Eqs.\ (\ref{PAD}) typically underestimate the nonlinear effects but they
yield a rough first estimate for $\gamma/\Omega$ less than about 0.25. More
interestingly, however, Eqs.\ (\ref{PAD}) show a general behavior which
is observed for the tunneling probability of the nonlinear problem as
well (see Fig.\ \ref{Fig1}). If $\gamma<0$,
$P_T^{(l)}$ is a monotonically decreasing function
of $\Omega^2/\varepsilon$ whereas for $\gamma>0$ it is oscillating.
This behavior is a consequence of the structure of the complex crossing
points of the quasienergy. For $\gamma>0$ three crossing points
contribute to $P_T^{(l)}$ thus causing interference effects whereas for
$\gamma<0$ there is only one. With a theory similar to the one of Ref.\
\cite{Dyk62} not available for nonlinear systems the study of this model
problem gives an idea of how these features in the behavior of $P_T$ may
come about and shows that they appear in a broader class of systems.
Numerical studies of the original Eqs.\ (\ref{NLZ}) indicate, however,
that at fixed $\gamma/\Omega > 0$
the minima of $P_T$ approximately occur at positions $k C \Omega^2/
\varepsilon$ with $k=1,2,\dots$, and $C$ a function of $\gamma/\Omega$.
This is in clear contrast to the behavior of Eqs.\ (\ref{PAD}),
furthermore, the first minimum appears at a much larger distance from
$\Omega^2/\varepsilon=0$ than indicated by these equations.
One might assume that it is possible to improve on the above results by
studying Eqs.\ (\ref{NLZ}) with $|a_+|^2$ replaced by the
exact stationary state population $N_{S,+}(t)$ as determined from
Eq.\ (\ref{sstat}). Unfortunately, the analytical theory of
Ref.\ \cite{Dyk62} is not applicable in this case as no complex
crossing points appear.
The numerically obtained tunneling probability has the same
qualitative features as discussed above; quantitatively, however,
the approximation to the exact $P_T$ is not as good
as one might intuitively expect (cf.\ Fig.\ \ref{Fig4}).

\section{Bloch band tunneling with Bose-Einstein condensates}

In this section the prospects of experimentally observing the nonlinear
effects discussed in Sec.\ III are examined with a focus on Bloch band
tunneling. To this end a one-dimensional simulation of Eq.\ (\ref{GPE})
for a sodium condensate is studied.
For the calculation, a condensate extension of
$l_z=160\mu$m and a period of the optical potential $\lambda=0.5\mu$m
are used. With $\Omega=10^4$s$^{-1}$ higher-order momentum
components are detuned by an amount of at least 40$\Omega$ from the
`$+/-$' modes when the resonance is crossed. The two conditions for the
applicability of approximations (\ref{expansion}), (\ref{NLZ}) which are
given in Sec.\ II are thus well fulfilled.
When the wave packet reaches the tunneling
resonance at $3\hbar k_L$ that follows the initial one at $\hbar k_L$
the two momentum components $+$ and $-$
have a detuning of 40$\Omega$, i.e., the tunneling process is well defined.
For $\Omega^2/\varepsilon\approx 1$ the acceleration is of the order
10ms$^{-2}$. The numerical simulations show that on the time scale of
passing through the resonance (less than 10$^{-2}$s in the examples)
the spreading of the initial wave packet as well as the spatial shift
between the two modes is small. The crucial parameter $\gamma$
takes on the value 10$^{-16}n_c$m$^3$s$^{-1}$ with $n_c$ the condensate
density. In our example a ratio of $\gamma/\Omega=1$ can be reached for a
BEC with $N=5\times 10^6$ atoms and a radius of 10$\mu$m. All the
parameter values given are well within the realm of currently feasible
experiments \cite{KetDurSta99}.

The numerical simulations of Eq.\ (\ref{GPE}) are performed with a
standard split-step algorithm (see, e.g., \cite{GarSuo95}). Thereby,
the initial state is
taken as the ground state of a harmonic potential yielding the required value of
$l_z$. The main results of these calculations are shown in Figs.\ \ref{Fig5}.
In these diagrams the fractional
population $P_+$ of the initial momentum component is displayed as a function
of $\tau=t m\varepsilon/4\hbar^2 k_L^2$, i.e., time measured in units of the
temporal distance between two resonance crossings. The point $\tau=0$
corresponds to zero detuning between the two relevant momentum modes.
Numerically, $P_+(t)$ is determined from the Fourier transform
of $\psi(z,t)$. Experimentally, it could be obtained by switching the
periodic potential off at $t$ and observing the subsequent condensate
evolution. In Fig.\ \ref{Fig5}(a) condensates with various values of
$\gamma/\Omega$ are started at rest with the acceleration chosen such that
the standard Landau-Zener formula (disregarding the atomic interactions)
predicts a tunneling probability of $0.01$. Having passed the resonance
$P_+$ quickly settles down to oscillations around a well-defined mean
value which may be regarded as the tunneling probability.
These mean values depend somewhat
sensitively on the initial conditions. For example, they may
change by up to 0.02 if the initial velocity is varied
by $\pm 0.01 \hbar k_L/m$. Nevertheless, they compare reasonably well to
the asymptotic tunneling probability
$P_T$ (indicated in Fig.\ \ref{Fig5} with filled circles $\bullet$)
obtained from Eqs.\ (\ref{NLZ}) for the corresponding parameter values.
Typically, the values for $P_T$ are smaller than the
\centerline{\psfig{figure=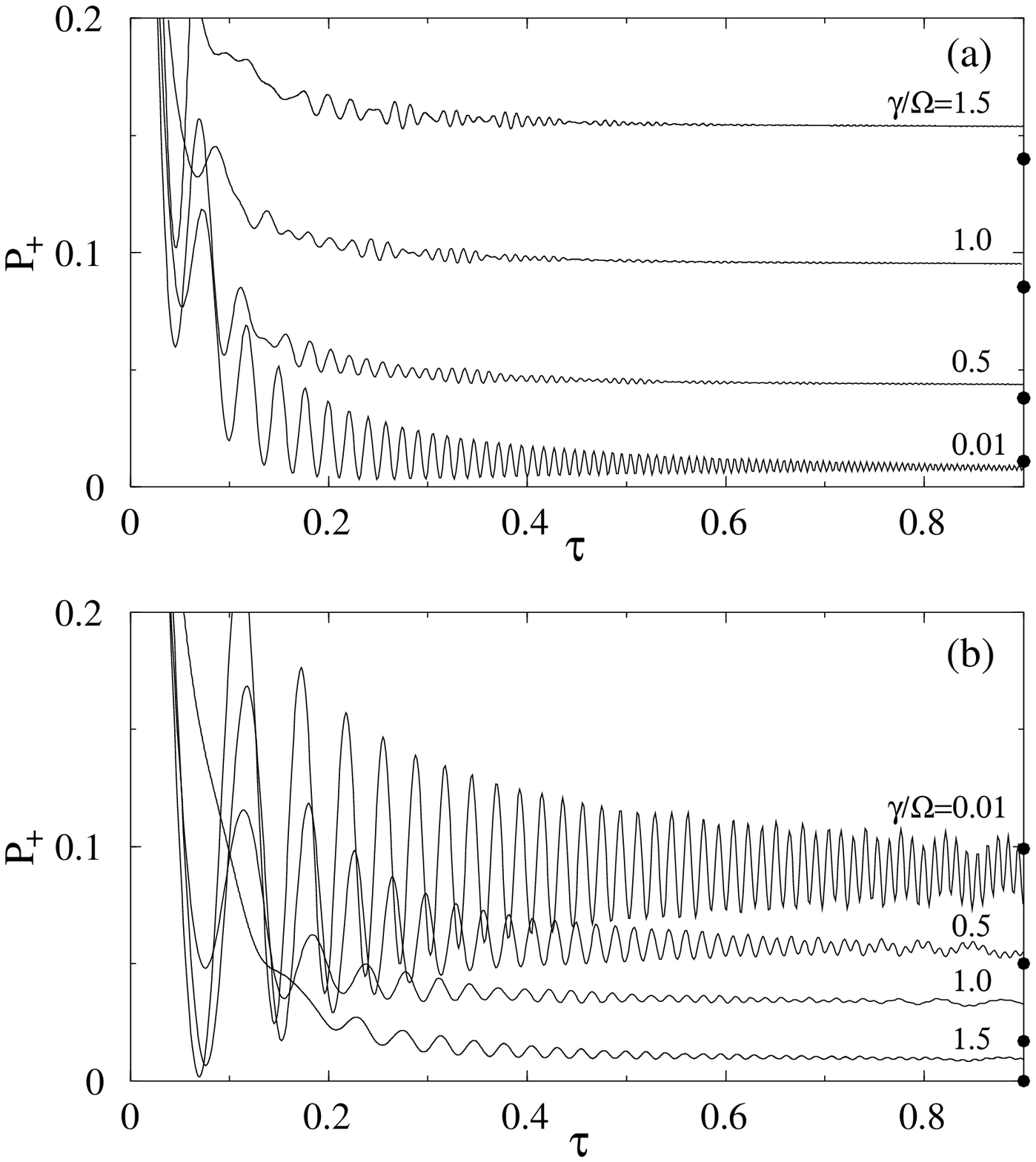,width=8.6cm,clip=}}
\begin{figure}
\caption{Fractional population $P_+$ of the initial momentum component
as a function of scaled time $\tau=t m\varepsilon/4\hbar^2 k_L^2$
for various values of $\gamma/\Omega$. In (a) the initial wave packet
velocity $v_i=0$ and the acceleration is chosen to obtain a tunneling
probability of $P_L=0.01$ according to the Landau-Zener formula in the absence
of nonlinear interactions. In (b) $v_i=-2\hbar/k_L m$ and $P_L=0.1$. The
predictions of Eqs.\ (\protect\ref{NLZ}) for $P_T$ at the respective parameter
values are indicated by filled circles $\bullet$.}
\label{Fig5}
\end{figure}
\noindent
results obtained from
the Gross-Pitaevskii equation by approximately 0.01$-$0.02.
Most importantly, however, it is seen that for $\gamma/\Omega=1.5$ the
tunneling probability is increased by almost an order of magnitude
compared to the case $\gamma/\Omega=0.01$. These results show that
$P_T$ is very sensitive to atomic interactions in experimentally
relevant situations. Figure \ref{Fig5}(b) illustrates that
one can also observe a significant decrease in the tunneling
probability. As discussed in Sec.\ III.A, this occurs if the condensate
starts from the high-energy state, i.e., in the present situation it
needs to be launched with a velocity less than $-\hbar k_L/m$.
In Fig.\ \ref{Fig5}(b) the initial velocity $-2\hbar
k_L/m$ was used which places the wave packet in the middle of two
tunneling resonances. The acceleration was chosen such
that without nonlinear interactions a tunneling probability of 0.1 is
expected. For smaller values of $\gamma/\Omega$ there is again good agreement
between the predictions of the two-mode model and the solution of the
Gross-Pitaevskii equation. At $\gamma/\Omega=2.0$, however, Eqs.\ (\ref{NLZ})
predict a tunneling probability of about 10$^{-4}$ whereas the simulation
yields a significantly higher value of $0.01$.

\section{Conclusion}

In this paper the influence of atomic interactions on time-dependent
tunneling processes of Bose-Einstein condensates was investigated. A
nonlinear Landau-Zener equation was derived which describes main aspects
of processes such as Bloch band tunneling, ground state population
transfer with variable external fields and condensate motion on coupled
potential surfaces. The tunneling probabilities predicted by this model
were discussed in detail, in particular, it was shown that for strong
enough nonlinearities a complete population transfer by adiabatic
following is impossible.
This behavior is a consequence of a nonlinear hysteresis effect.
To assess the reliability of the nonlinear Landau-Zener model a
comparison to simulations of the Gross-Pitaevskii equation was made.

In actual experiments
it is only possible to determine the tunneling
probability for a certain portion of the $(\Omega^2/\varepsilon,
\gamma/\Omega)$ parameter space. For example, the acceleration cannot be
made arbitrarily small so that the asymptotic behavior of the tunneling
probability shown in Fig.\ \ref{Fig2}(b) may never be verified directly.
Furthermore, tunneling probabilities close to 0 or 1 are very difficult
to measure accurately due to the long-time oscillations of the mode
populations. Nevertheless, as shown in Sec.\ IV, drastic nonlinear
effects already appear in the experimentally accessible parameter
space and they are appropriately described with the simple model of
Eqs.\ (\ref{NLZ}). Its detailed study is thus well justified.

However, to obtain a full understanding of time-dependent tunneling it is
necessary to expand the investigation beyond the limits of applicability
of Eqs.\ (\ref{NLZ}). In the context of Bloch band tunneling new
features in the system behavior may arise, for example, if the extension
of the condensate becomes comparable to the period of the optical potential.
Another
potentially interesting question concerns the study of quantum mechanical
effects beyond the mean field description that was applied here.

This work was supported by the United Kingdom Engineering and Physical
Sciences Research Council.

\section*{Appendix: Derivation of Eqs.\ (\protect\ref{PAD})}

The complex crossing points $t_k$ are obtained as solutions of
\begin{equation}\label{poly}
Q^2(t)=\frac 1 4 \left(\Delta -\frac{\gamma
\Delta}{2\sqrt{\Delta^2/4+\Omega^2}}\right) ^2+\Omega^2=0,
\end{equation}
with $\Delta=\varepsilon t$, which can be converted into a quartic
equation. If $\gamma < 0$ there is only one root in the upper complex
half plane which is given by
$$\Delta_1
=2i\Omega[1-(|\gamma|/\Omega)^{2/3}/2+\rm O(|\gamma|^{4/3})].$$
The integral ${\cal D}_1$ is approximated by expanding the integrand
as $$Q(t)\approx\sqrt{\Delta^2/4+\Omega^2}-
\gamma \Delta^2/[8(\Delta^2/4+\Omega^2)].$$
The integration can then be carried out analytically and the second of
Eqs.\ (\ref{PAD}) is obtained with $\Gamma_1=-1$.

In the case $\gamma > 0$ Eq.\ (\ref{poly}) has three solutions in the upper
complex half plane. They are given by 
\begin{eqnarray*}
\Delta_{1,2}&=&2i\Omega\{1+e^{\pm
i\pi/3}(\gamma/\Omega)^{2/3}/2+\rm O(\gamma^{4/3})\}, \\
\Delta_3 &=&2i\Omega\left[1-\frac 1 {32} \frac{\gamma^2}{\Omega^2} + \rm O
(\gamma^4)\right]
\end{eqnarray*}
with $\Gamma_{1,2}=-1$ and $\Gamma_3=1$. The integrals ${\cal D}_{1,2}$
are determined as in the case $\gamma<0$. The integral ${\cal D}_3$
cannot be dealt with in this way. However, a numerical analysis shows
that for $\gamma/\Omega<1$ one can set to a good degree of approximation
${\cal D}_3=i \rm{Im}({\cal D}_{1,2})$. This approach yields the first
of Eqs. (\ref{PAD}). The comparison of these equations to the numerically
determined $P_T^{(l)}$ confirms the accuracy of the approximation for
$|\gamma|/\Omega<1$.

\end{document}